# Dual-band electro-optically steerable antenna

Dmytro Vovchuk, Anna Mikhailovskaya, Dmitry Dobrykh, and Pavel Ginzburg

*Abstract*— The ability to obtain dynamic control over an antenna radiation pattern is one of the main functions, desired in a vast range of applications, including wireless communications, radars, and many others. Widely used approaches include mechanical scanning with antenna apertures and phase switching in arrays. Both of those realizations have severe limitations, related to scanning speeds and implementation costs. Here we demonstrate a solution, where the antenna pattern is switched with optical signals. The system encompasses an active element, surrounded by a set of cylindrically arranged passive dipolar directors, functionalized with tunable impedances. The control circuit is realized as a bipolar transistor, driven by a photodiode. Light illumination in this case serves as a trigger, capable of either closing or opening the transistor, switching the impedance between two values. Following this approach, a compact half-a-wavelength footprint antenna, capable to switch between 6 dBi directional patterns within a few milliseconds' latency was demonstrated. The developed light activation approach allows constructing devices with multiple almost non-interacting degrees of freedom, as brunched feeding network is not required. The capability of MHz and faster switching between multiple electromagnetic degrees of freedom open pathways to new wireless applications, where fast beam steering and beamforming performances are required.

*Index Terms*—steerable antenna, electro-optical control, dual-band, compact antenna, latency.

## I. INTRODUCTION

THE ABILITY to control the radiation pattern with high accuracy allows for establishing efficient point-to-point communication, where one or more participants can change their locations during the process. A radar, tracking a moving target in both azimuth and elevation, is one notable example. Recently, the automotive industry raised a demand for high-resolution short-range radar-based imaging systems, where high-quality fast scanning small aperture antennas are essential components [1]–[3]. Another realm is 5G communications, where beamforming with millisecond-scale latency is the enabling technology to support fast-speed broadband wireless communication [4], [5]. In all the beforehand mentioned applications, antenna devices are subject to engineering tradeoffs where high scanning speed and low cost are contradictory requirements. There are several traditional approaches to beam steering. The first one is a mechanical scan, where a motor controls the angular position of a highly directive antenna. This technique is frequently used for implementing marine and airport tracking radars, where scanning speeds are not the main factor to consider. Another approach to beam steering is based on antenna phased arrays. Here multiple elements are phased-locked and radiate simultaneously. While this architecture allows achieving fast all-electronic scanning, the realization of high-quality and directive beams requires employing tens or even hundreds of phase-shifting elements. This approach is used e.g., in airborne applications, where the speed and scan quality requirements predominate over-involved costs of realizations. Recently, several approaches, complementary to traditional phased arrays have been proposed and demonstrated. The ability to tailor and control the laws of refraction with the help of artificially structured media (metamaterials [6]–[9]) opened a range of new capabilities in beam shaping and control. Carefully designed surfaces (metasurfaces) can provide capabilities to tailor properties of transmitted and reflected waves [6], [10]–[12]. While many metasurface studies concentrate on static configurations (e.g., [13]–[15]), introducing fast real-time tunability is the demanded feature. Several realizations of dynamically reconfigurable metasurfaces and metasurface-based antennas have been demonstrated (e.g. [11], [16]–[20]). The key underlining concept is typically based on controlling individual resonant elements within an array with electronics. For example, tunable capacitance allows shifting resonant responses of individual elements, and as the result, either amplitude or phase switchable screens are achieved [11], [21], [22]. While this type of realization does not rely on expensive phase shifters, it still requires using numerous (yet simple and cheap) electronic elements, and, even more critically, a branched set of wires to drive them. While reflect array configurations allow hiding wires behind a ground plane [23], [24], electric circuitry can significantly affect electromagnetic performances in other realizations. For example, a mesh of thin wires with subwavelength spacing will have a predominating undesired electromagnetic response. A probable solution to this problem has been demonstrated in the case of volumetric metamaterial-based scatterers [25]. It relies on driving individual meta-atoms with light. Light and light guiding materials do not interact with cm and mm waves, which enables uncoupling of these two phenomena in the design. The interaction happens directly within an individual antenna

(Corresponding author: Dmytro Vovchuk e-mail: dimavovchuk@gmail.com).
Dmytro Vovchuk, Anna Mikhailovskaya, Dmitry Dobrykh, Pavel Ginzburg School of Electrical Engineering, Tel Aviv University, Ramat Aviv, Tel Aviv, 69978, Israel (e-mail: dimavovchuk@gmail.com, pginzburg@tauex.tau.ac.il ).

Dmytro Vovchuk Department of Radio Engineering and Information Security, Yuriy Fedkovych Fernivtsi National University, Chernivtsi, 58012, Ukraine (e-mail: dimavovchuk@gmail.com)
Dmytro Vovchuk and Anna Mikhailovskaya contributed equally to this work.

element, where optical energy is rectified within a photoelement to drive electronics. Here we develop the concept of electro-optically driven beamforming, which allows fast manipulation over radiation patterns by arranging arrays of auxiliary optically switchable reflectors and directors around a radiating element. The optical link allows for obtaining both high switching speeds and modularity, i.e., almost any radiating element can be granted with scanning capabilities, as the constraints, related to a wired feeding network are relaxed.

The manuscript is organized as follows – the design and implementation of a single reflector are introduced first and then followed by its integration into an antenna device. Beam steering performances are assessed next along with investigating of other antenna characteristics. Measurements of the beam steering rates are demonstrated next. The capability to grant steering capabilities for several commercial and custom-made antennas radiating elements is discussed before the Conclusion.

## II. ELECTRO-OPTICALLY DRIVEN ELEMENT

Quite a few designs of directive antennas are based on interference phenomena between several elements. A representative example here is Yagi-Uda antenna, where a set of passive elements – reflectors and directors, are responsible for a narrow beam formation. Each of them introduces a different phase lag, which is tuned by controlling lengths of elements within the architecture. While physical size of a resonant element cannot be controlled dynamically on a reasonably fast timescale, electric length can be governed by introducing a tunable lumped element. As the first step, we will demonstrate a design of wirelessly tunable single element, which will be subsequently integrated within a beam steering array. Two states – 'on' and 'off' correspond to either presence or absence of the illumination. Our basic component is a half-wavelength element ($\lambda/2$), formed by a pair of $\lambda/4$-lenght wires with a gap in between (Fig. 1(a)). The driving circuit consists of two photodiodes (BPW34) and a bipolar transistor (BFU730F115 NPN-type BJT) as in Fig. 1(b). Two photoelements are used to elevate the voltage drop to open the transistor. If the illumination power on the circuit is insufficient, the element acts as a cut. After passing a threshold, the diode become a shortage.

Figure 1(c) demonstrates the forward scattering spectra of the system at its two states. Wire dimensions are length l = 72 mm and radius r = 0.5 mm. The gap at the middle is 1mm. Those parameters were tuned to make the device complying with IEEE 802.11 communication standard (in terms of radiation bands). It is worth mentioning that the transistor impedance is also considered for both open and short operation states. Fig. 1(c) demonstrates the capability to tune the scattering peak from 2.2GHz to 1.9 and vice versa upon light illumination. Full-wave numerical analysis, including an introduction of lumped elements, was done with CST Microwave Studio. The surface current distribution on the element strongly depends on the light state (insets to Fig. 1(c)), demonstrating the switching between dipolar and quadrupolar operation modes. 0.5 pF of the lumped element C was found to provide a reliable model to switching for the state 'off' and a solid $\lambda/2$ wire – for the state 'on'. Slight differences between numerical and experimental data come from nonvanishing formfactors of active elements, which were not considered in simulations.

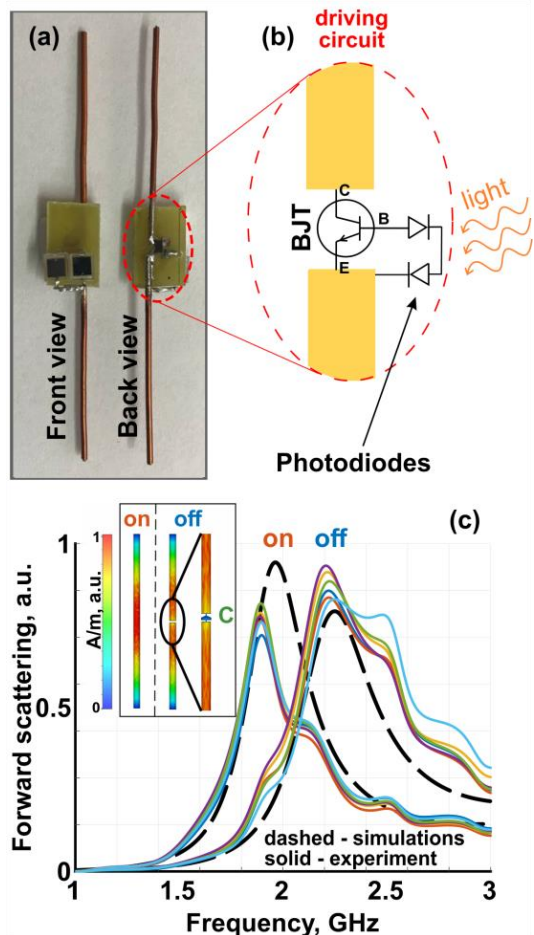

Fig. 1. Optically-switchable passive element – photograph (a) and the schematics (b) of the photo-activated driving circuit (BJT – bipolar junction transistor, C – collector, B – base and E – emitter). (c) Numerical analysis and experimental forward scattering spectra of the device at light 'on' and 'off' states. Color lines – responses of individual elements. Inset – current distributions along the elements (numerical results).

A choice of elements for implementing the driving circuit worth a discussion. Among possible architectures (i) varactor + photodiode; (ii) PIN-photodiode and (iii) phototransistor or transistor + photodiode can be considered. While varactors are commonly used in related designs [10], [11], those are not the best candidates for the current implementation as they demand quite high voltage to provide a pF-scale capacitance tunability. 0.7 V for Si and 0.35 V for Ge implementations are requited. Phototransistors are typically designed for low-frequency applications, e.g., fire protection or motion detection. Therefore, we will investigate a combination of a low-cost high-frequency BFU730F115 npn-type BJT and BPW34 photodiodes. The photodiode's anode is connected to the transistor's base and cathode to the emitter (Fig. 1(b)). The collector and emitter of the transistor are the outputs of the driving circuit and are soldered to the $\lambda/4$ wires. This arrangement allows shifting the scattering resonance to higher frequencies.

## III. Optically Steerable Antenna

After designing single elements, those will be assembled to form a larger-scale system, which aims on providing beam steering capabilities. Six passive director elements were chosen to form the geometry. This number, being found beneficial to optimize wire bundle scatterers [26]–[29], was chosen as a tradeoff between design simplicity and functionality. While this configuration fits demands of 6-sector 4G wireless network, it can be further tuned per application, i.e., the number of scanning lobes can be increased, and various 5G communication protocols can be implemented.

The antenna consists of seven elements in overall: one active (marked with '#') placed exactly at the center and six passives (1-6) are equidistantly placed on an imaginary cylindrical surface (Fig. 2). A broadband monopole antenna (W1096), covering the investigated frequency range and providing rather flat frequency response, was chosen as a feed [30]. This commercial element can be replaced by a custom-made monopole, tuned per frequency. Before assembling the structure, each of six passive elements was calibrated to provide the identical response (as in Fig. 1(c)). Here both scattering parameters and optical activation power are adjusted. Each individual element was checked separately by performing a forward scattering experiment. As the element acts as a dipole, this parameter almost completely characterizes its response. The manual adjustment was done by cutting the wire's length. It is also worth noting that nominals of lumped elements can vary from item to item. Hence, an individual calibration is required. Fig. 1(c) demonstrates the calibration curves, the averaged parameters of which was used as in antenna modeling.

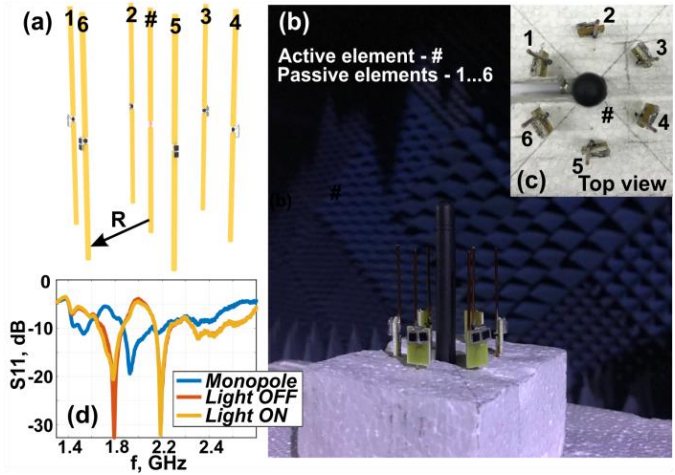

Fig. 2. (a) Schematic layout and (b) photograph of the optically steerable antenna. On the insets (c) photograph of the top view. (d) $S_{11}$ parameters of antennas – standalone monopole, steering antenna with light 'on and 'off', as in the legends.

Without a light activation, all six passive elements are identical and, as a result, the radiation pattern has no directivity in-plane (end-fire). To break the symmetry, several elements can be triggered with light. For an initial approximate analysis, the elements can be considered as present for 2.2GHz wave if the light is "on" and absent if there is no direct illumination on them. For 1.8GHz the scenario is reversed. As a result, several elements form a directive pattern. A more accurate analysis suggests considering impact of non-resonant inactivated elements. This was done numerically, and the system parameters were additionally optimized. The optimization is applied to maximize directivity and gain of the antenna, constraining its overall size [31]. While a directivity in a Yagi-Uda antenna relies on interference phenomena between several directors and reflectors, the proposed realization involves multipolar interaction and near-field coupling between elements [26]–[28]. The radius of the imaginary cylindrical surface (taking into account the cylinder radius R = 20 mm), containing optically switchable passive elements, was chosen to be 41 mm ≈ 0.26λ. 1.8 and 2.2GHz were chosen quite arbitrary within the wireless band and can be tuned per a specific application.

Fig. 3 summarizes the patterns, obtained both numerically and experimentally at an anechoic camber. '1' and '0' in the figure captions indicate whenever the element was illuminated or not, respectively. Antenna matching conditions ($S_{11}$ parameters) appear in Fig. 2(d). While the initial design was made for a single-element activation (Fig. 3a-d), different combinations can be considered as well. Theoretically the system has 2N independent degrees of freedom, where N is the number of elements. Potentially, 2N antenna patterns can be achieved, nevertheless not all of them can be considered as practically relevant. Several reports have demonstrated N patterns with N tunable elements [24], [32], [33]. While our structure was not designed to maximize the number of patterns, we found that activating pairs of adjacent elements leads to formation of directional beams, shifted by 30° in respect to the single-element case (Fig. 3e-h). As a result, we have demonstrated 12 directional beams, i.e., 2N useful patterns. Furthermore, the device shows a dual band performance – both 1.8 and 2.2 GHz with a 10% fractional bandwidth. Activating other combination of elements didn't lead to formation of patterns with reasonable directivity.

Directivity (D) and gain (G) of the antenna will be characterized next. As the pattern is formed primarily in-plane, the following relation will be used to process the experimental data [34]:

$$D(\varphi, \theta = \text{const}) = \frac{P_{max}}{\frac{1}{2\pi}\int_0^{2\pi} P(\varphi)d\varphi}, \quad (1)$$

where $P_{max}$ is the maximal radiated power of the antenna. The assessment is made for a constant elevation angle ($\theta = 0$) and for the entire $2\pi$ of the azimuth $\varphi$. The realized gain $G_{Tx}$ is extracted by comparing the device with an etalon antenna (IDPH-2018 S/N-0807202 horn) with a known gain $G_{Rx}$. Eq. 2 is used for the analysis [34].

$$G_{Tx} = \left(\frac{4\pi a}{\lambda}\right)^2 \frac{P_{Rx}}{P_{Tx}} \frac{1}{G_{Rx}}, \quad (2)$$

where 'a' is the distance between the apertures of the transmit Tx and the receive Rx antenas, λ is the operational wavelength and $P_{Rx}/P_{Tx} = |S_{21}|^2$ is the power transmission coefficient.

To assess the switching parameter, we calculated the differential gain values between the 'on' and 'off' states ($G_{on}$ and $G_{off}$), as following:

$$G_{diff} = G_{on} - G_{off}. \quad (3)$$

The results are summarized in Table I. The numerical results on directivity are presented for the 2D (φ,θ=0) and 3D (φ,θ) cases, while the experiments are shown only for 2D case. One can see the difference between the directivity of numerical and experimental values, especially at 2.2 GHz. The results can be assessed by comparing patterns in Fig. 3. The most pronounced difference was found for the data on panels (c) and (d). A significant back lobe, being predicted numerically (imperfect optimization), was not found in the measurements. The opposite behavior was found for the two-element illumination at 2.2 GHz – here back lobes were found in the experiment, while the numerical prediction suggests rather minor back radiation. The reason for this can be several-fold: (i) imperfection in elements, affecting the interference phenomena and (ii) a parasitic illumination due to the ambient illumination and the pollution from nearby light sources – the driving LED (as will be discussed hereinafter). Nevertheless, the back lobe suppression effect is not dramatic. (iii) Nevertheless, the feeding monopole connector has an orientation, perpendicular to the antenna axis, it breaks the symmetry between different radiation patterns (e.g., yellow, and purple lines in Fig. 3).

It is worth mentioning that the system cannot perform an independent simultaneous beam steering at two different frequency bands, as the same photodiodes are in use.

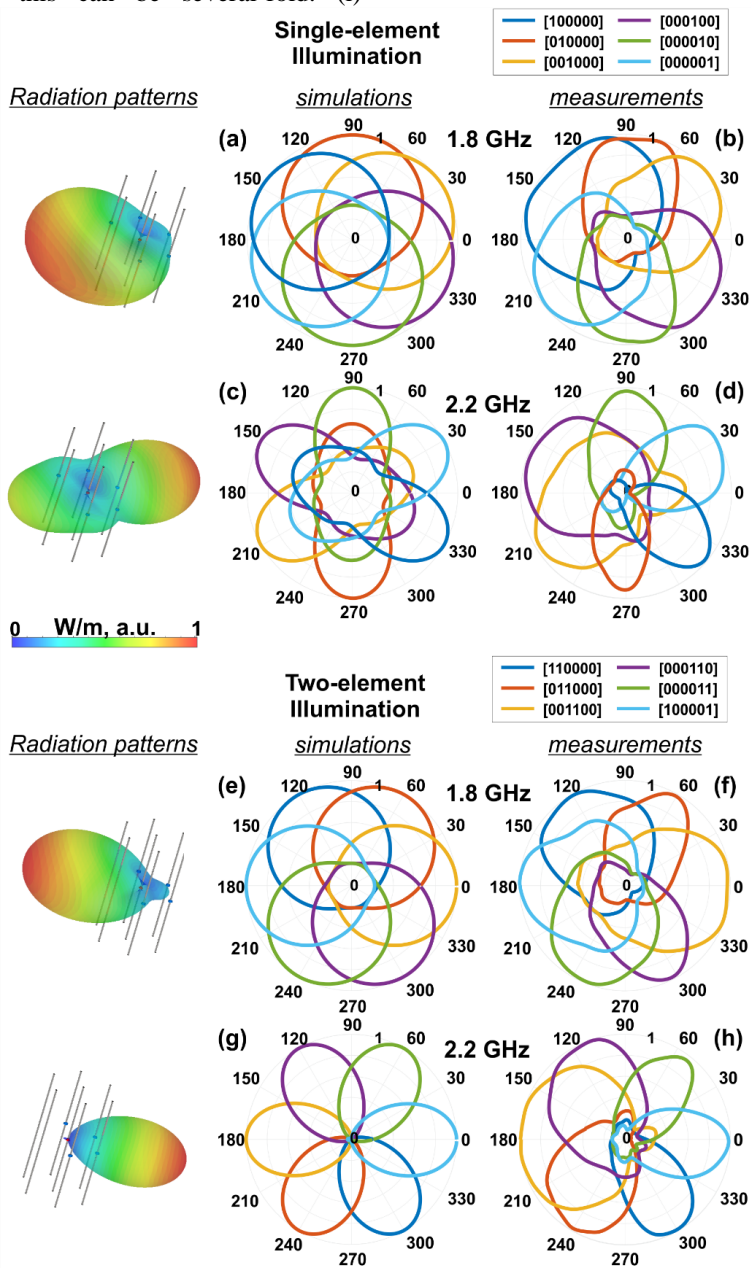

Fig. 3. Radiation patterns – numerical and experimental results. Single (a-d) and double-element (e-h) illumination at the frequencies 1.8 (director case) and 2.2 GHz (reflector case). Antenna 3D radiation patterns (numerical results) are in left insets.

TABLE I
The directivity $D$ and differential gain $G_{diff}$.

|  |  | f, GHz | Numerical | | Experimental |
|---|---|---|---|---|---|
|  |  |  | 2D | 3D | 2D |
| Single-element illumination | D, dBi | 1.8 | 2.68 | 5.21 | 3.31 |
|  |  | 2.2 | 2.48 | 5.17 | 4.11 |
|  | $G_{diff}$, dBi | 1.8 |  | 2.06 | 2.65 |
|  |  | 2.2 |  | 5.68 | 5.56 |
| Two-element illumination | D, dBi | 1.8 | 3.36 | 6.01 | 3.37 |
|  |  | 2.2 | 6.17 | 9.27 | 3.85 |
|  | $G_{diff}$, dBi | 1.8 |  | 2.49 | 2.2 |
|  |  | 2.2 |  | 7.25 | 4.62 |

Free-space illumination of photodiodes requires an extra-consideration. The first factor is an ambient radiation, which can accidentally bring the system to a threshold. For an assessment, we compared chamber conditions with an office space and outdoors (direct summer sunlight). In last two cases, a light concealment arrangement is required to maintain the correct operation of the device. The second factor is undesired light from a nearby illuminated element. The distance between the LED and the photodiode is 1cm (inset to Fig. 4(a), thus the light leakage was found to play no role. In both cases the voltage on the diode was measured and compared with 0.7V threshold. It is worth noting that introducing integrated optics arrangements (e.g., waveguiding devices) are capable to solve issues of the undesired overexposure to light.

One of the main advantages of the proposed design is its potentially fast switching rates. 5G standards demand latencies as a small as a milli-second. It implies having capabilities of sub-MHz beam steering rates. To assess this parameter, the following setup have been constructed – a signal from a high-frequency generator (N5173B EXG X-Series Microwave Analog Signal Generator) is split via ZX10-2-852-S+ Splitter into two channels: the first feeds the active element of the antenna and the second provides the synchronization signal and feeds the local oscillator (LO) input of a mixer ZX05-C24-S+ at the receiver (Fig. 4(a)). The LF pulse sequences generator (81160A Pulse Function Arbitrary Noise Generator) feeds a LED SMD5630, which is located close to the antenna photodiodes and performs the on/off-switching with a period T = 1 ms. 50% duty cycle ($\tau$) was chosen. The receiver includes Rx antenna, feeding the RF input of the mixer. The output, after a low-pass filter (LPF) BLP-100-75+, is displaced on a scope. The digitalized scope's output allows investigating switching properties of the device (antenna under test – AUT). The results show that $f_0 = 1/T$ at 1 kHz can be obtained (Fig. 4(b)). To determine rise ($t_r$) and fall ($t_f$) times, the received signal was smoothed and fitted with a sine series (Fig. 4(c)). The extracted rise and fall times for the system are ~ 0.1 ms.

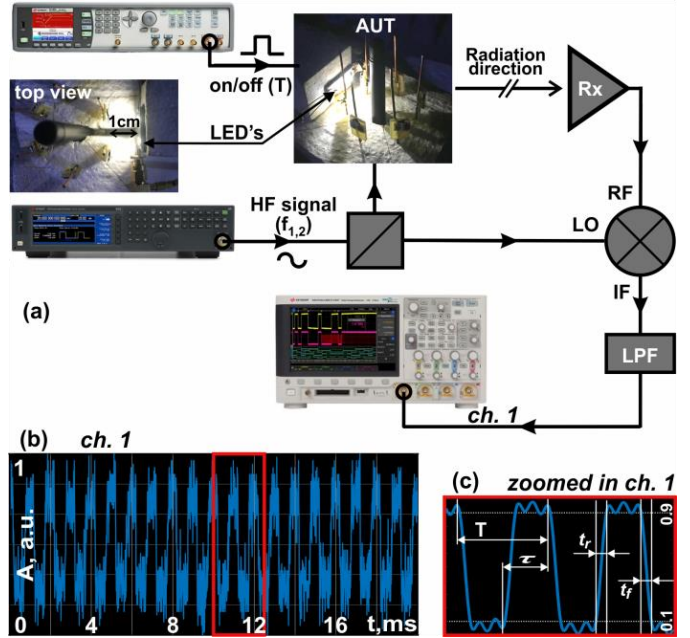

Fig. 4. (a) Schematics of the setup for measuring the switching rate. (b) Zoomed IF signal on the scope. (c) Post-processed signal - period T = 1 ms (50% duty cycle for $\tau$ = 0.5 ms), rise $t_r$, and fall $t_f$ time.

IV. BEAM STEERING WITH OTHER ANTENNAS

To demonstrate the flexibility of the proposed method, 3 different antennas have been considered, namely the commercial monopole from the previous studies, symmetric dipole antenna and a monopole above a ground plane (panels a, d, and f in Fig. 5, respectively). Each of those has an omni-directional pattern in-plane. Two switching elements has been used do demonstrate the concept. As the structures have reflection symmetry, only one directional pattern per frequency was demonstrated. Yellow and green lines correspond to 2.2 and 1.8 GHz, respectively. Illuminating one side of the structure leads to a creation of directional patterns, which are oppositely oriented for both of those frequencies. Switching between the illumination side will case the flip in the patters. The commercial monopole antenna has slightly better performances owing as it underwent a significant optimization by the vendor. The dipole demonstrates less directive pattern at 1.8GHz owing to the frequency-dependent balun. This aspect does not affect the monopole configuration, which also demonstrates good switching capabilities.

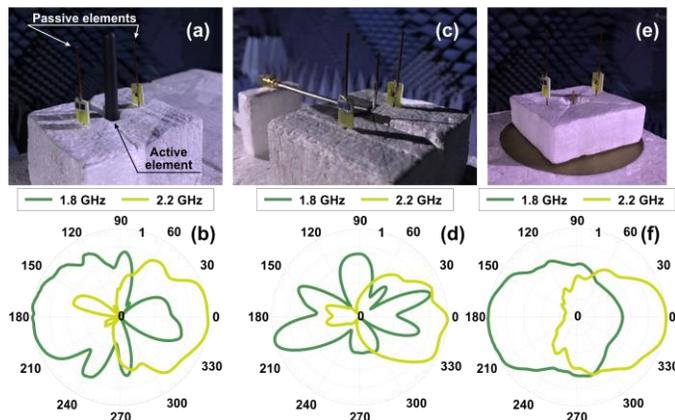

Fig. 5. The concept of granting a radiating element with beam steering capabilities. (a), (c), and (e) – photographs of antenna devices. (b), (d), and (f) - experimentally obtained (in-plane) radiation patterns. Switching between 2 sectors has been considered.

## V. Conclusion

A scanning antenna with optical control is demonstrated experimentally. The device consists of six passive resonators, arranged around the feed. Electromagnetic properties of passive elements, serving as either directors or reflectors, are tuned with light. The driving circuit, containing photodiodes and bipolar transistor, is activated remotely with light. This approach allows tuning electromagnetic properties of the system without a need of a brunched network of metal wires. The demonstrated design provides steering capabilities of directional beams with ~5 dBi of the directivity and 6 dBi of the differential gain with a switching rate around at sub-MHz rate. The demonstrated antenna belongs to the class of compact ($2r/\lambda \approx 0.5$-$0.6$, where r is the radius of an imaginary sphere that surrounds the whole antenna [31], [35]) low-cost devices (the active element + six passive elements with driving circuits cost around 20$). Furthermore, it was shown to provide a dual-band operation at frequencies, relevant to wireless communications. Further optimization of the electromagnetic design and introduction of fast elements (transistors and fast photodiodes) can elevate the switching rates towards MHz and higher opening pathways to new applications, where fast beam steering and beamforming performances are required (e.g., radars and 5G). Frequency bands in 5G protocols are quite broad and utilized per application, though a capability of fast beam control remains essential. Light activation approach allows constructing devices with multiple almost non-interacting degrees of freedom, as brunched feeding network is not required and, in principle, almost any radiating element can be granted with beam steering capabilities.

ACKNOWLEDGEMENTS

The work was supported by ERC POC, grant 101061890 "DeepSight".

References

[1] S. M. Patole, M. Torlak, D. Wang, and M. Ali, "Automotive Radars: A review of signal processing techniques," *IEEE Signal Process. Mag.*, vol. 34, no. 2, pp. 22–35, Mar. 2017, doi: 10.1109/MSP.2016.2628914.

[2] A. Asensio-López et al., "High range-resolution radar scheme for imaging with tunable distance limits," *Electron. Lett.*, vol. 40, no. 17, pp. 1085–1086, Aug. 2004, doi: 10.1049/EL:20045552.

[3] MathWorks, "5G Development with MATLAB," *MathWorks*, 2017, [Online]. Available: https://uk.mathworks.com/content/dam/mathworks/tag-team/Objects/5/5G_ebook.pdf

[4] "Intel 5G Standards and Spectrum." https://www.intel.com/content/www/us/en/wireless-network/5g-technology/standards-and-spectrum.html (accessed Jun. 21, 2022).

[5] R. Ford, M. Zhang, M. Mezzavilla, S. Dutta, S. Rangan, and M. Zorzi, "Achieving Ultra-Low Latency in 5G Millimeter Wave Cellular Networks," *IEEE Commun. Mag.*, vol. 55, no. 3, pp. 196–203, Mar. 2017, doi: 10.1109/MCOM.2017.1600407CM.

[6] N. Engheta and R. Ziolkowski, *Electromagnetic Metamaterials: Physics and Engineering Explorations*. 2006. doi: 10 0-471-76102-8.

[7] D. Filonov, A. Shmidt, A. Boag, and P. Ginzburg, "Artificial localized magnon resonances in subwavelength meta-particles," *Appl. Phys. Lett.*, vol. 113, no. 12, p. 123505, Sep. 2018, doi: 10.1063/1.5047445.

[8] V. S. Asadchy, M. Albooyeh, S. N. Tcvetkova, A. Díaz-Rubio, Y. Ra'Di, and S. A. Tretyakov, "Perfect control of reflection and refraction using spatially dispersive metasurfaces," *Phys. Rev. B*, vol. 94, no. 7, p. 075142, Aug. 2016, doi: 10.1103/PHYSREVB.94.075142/FIGURES/8/MEDIUM.

[9] F. Capolino, "Applications of Metamaterials," *Appl. Metamaterials*, Dec. 2017, doi: 10.1201/9781420054248/APPLICATIONS-METAMATERIALS-FILIPPO-CAPOLINO.

[10] N. I. Zheludev and Y. S. Kivshar, "From metamaterials to metadevices," *Nature Materials*, vol. 11, no. 11. Nature Publishing Group, pp. 917–924, Oct. 23, 2012. doi: 10.1038/nmat3431.

[11] V. Kozlov, D. Vovchuk, and P. Ginzburg, "Broadband radar invisibility with time-dependent metasurfaces," *Sci. Reports 2021 111*, vol. 11, no. 1, pp. 1–11, Jul. 2021, doi: 10.1038/s41598-021-93600-2.

[12] M. Faenzi et al., "Metasurface Antennas: New Models, Applications and Realizations," *Sci. Reports 2019 91*, vol. 9, no. 1, pp. 1–14, Jul. 2019, doi: 10.1038/s41598-019-46522-z.

[13] H. Markovich, D. Filonov, I. Shishkin, and P. Ginzburg, "Bifocal Fresnel Lens Based on the Polarization-Sensitive Metasurface," *IEEE Trans. Antennas Propag.*, vol. 66, no. 5, pp. 2650–2654, May 2018, doi: 10.1109/TAP.2018.2811717.

[14] V. Kozlov, D. Filonov, A. S. Shalin, B. Z. Steinberg, and P. Ginzburg, "Asymmetric backscattering from the hybrid magneto-electric meta particle," *Appl. Phys. Lett.*, vol. 109, no. 20, p. 203503, Nov. 2016,

doi: 10.1063/1.4967238.
[15] D. Filonov, V. Kozlov, A. Shmidt, B. Z. Steinberg, and P. Ginzburg, "Resonant metasurface with tunable asymmetric reflection," *Appl. Phys. Lett.*, vol. 113, no. 9, p. 094103, Aug. 2018, doi: 10.1063/1.5046948.
[16] H.-X. Xu *et al.*, "Tunable microwave metasurfaces for high-performance operations: dispersion compensation and dynamical switch," *Sci. Rep.*, vol. 6, no. 1, p. 38255, Dec. 2016, doi: 10.1038/srep38255.
[17] D. F. Sievenpiper, J. H. Schaffner, H. J. Song, R. Y. Loo, and G. Tangonan, "Two-dimensional beam steering using an electrically tunable impedance surface," *IEEE Trans. Antennas Propag.*, vol. 51, no. 10, pp. 2713–2722, Oct. 2003, doi: 10.1109/TAP.2003.817558.
[18] W. Yang, L. Gu, W. Che, Q. Meng, Q. Xue, and C. Wan, "A novel steerable dual-beam metasurface antenna based on controllable feeding mechanism," *IEEE Trans. Antennas Propag.*, vol. 67, no. 2, pp. 784–793, Feb. 2019, doi: 10.1109/TAP.2018.2880089.
[19] X. Wang and S. Tretyakov, "From Tunable and Reconfigurable to Space-Time Modulated Multifunctional Metasurfaces," *2021 IEEE Int. Symp. Antennas Propag. North Am. Radio Sci. Meet. APS/URSI 2021 - Proc.*, pp. 1361–1362, 2021, doi: 10.1109/APS/URSI47566.2021.9704202.
[20] M. Di Renzo *et al.*, "Smart Radio Environments Empowered by Reconfigurable Intelligent Surfaces: How It Works, State of Research, and the Road Ahead," *IEEE J. Sel. Areas Commun.*, vol. 38, no. 11, pp. 2450–2525, Nov. 2020, doi: 10.1109/JSAC.2020.3007211.
[21] D. Ramaccia, D. L. Sounas, A. Alu, A. Toscano, and F. Bilotti, "Phase-Induced Frequency Conversion and Doppler Effect with Time-Modulated Metasurfaces," *IEEE Trans. Antennas Propag.*, vol. 68, no. 3, pp. 1607–1617, Mar. 2020, doi: 10.1109/TAP.2019.2952469.
[22] J. J. Luther, S. Ebadi, and X. Gong, "A microstrip patch electronically steerable parasitic array radiator (ESPAR) antenna with reactance-tuned coupling and maintained resonance," *IEEE Trans. Antennas Propag.*, vol. 60, no. 4, pp. 1803–1813, Apr. 2012, doi: 10.1109/TAP.2012.2186265.
[23] C. Sun, A. Hirata, T. Ohira, and N. C. Karmakar, "Fast beamforming of electronically steerable parasitic array radiator antennas: Theory and experiment," *IEEE Trans. Antennas Propag.*, vol. 52, no. 7, pp. 1819–1832, 2004, doi: 10.1109/TAP.2004.831314.
[24] M. Rzymowski, D. Duraj, L. Kulas, K. Nyka, and P. Woznica, "UHF ESPAR antenna for simple angle of arrival estimation in UHF RFID applications," *2016 21st Int. Conf. Microwave, Radar Wirel. Commun. MIKON 2016*, Jun. 2016, doi: 10.1109/MIKON.2016.7491984.
[25] D. Dobrykh, A. Mikhailovskaya, P. Ginzburg, and D. Filonov, "4D Optically Reconfigurable Volumetric Metamaterials," *Phys. status solidi – Rapid Res. Lett.*, vol. 14, no. 8, p. 2000159, Aug. 2020, doi: 10.1002/PSSR.202000159.
[26] D. Vovchuk, S. Kosulnikov, R. E. Noskov, and P. Ginzburg, "Wire resonator as a broadband Huygens superscatterer," *Phys. Rev. B*, vol. 102, no. 9, 2020, doi: 10.1103/PhysRevB.102.094304.
[27] S. Kosulnikov *et al.*, "Circular wire-bundle superscatterer," *J. Quant. Spectrosc. Radiat. Transf.*, vol. 279, p. 108065, Mar. 2022, doi: 10.1016/J.JQSRT.2022.108065.
[28] K. Grotov *et al.*, "Genetically Designed Wire Bundle Super-Scatterers," *IEEE Trans. Antennas Propag.*, pp. 1–1, 2022, doi: 10.1109/TAP.2022.3177531.
[29] R. W. P. King, G. J. Fikioris, and R. B. Mack, *Cylindrical Antennas and Arrays*. Cambridge, 2002.
[30] "W1095X Datasheet(PDF) - Pulse A Technitrol Company." https://www.alldatasheet.com/datasheet-pdf/pdf/1320661/PULSE/W1095X.html (accessed Jun. 21, 2022).
[31] M. Pigeon, C. Delaveaud, L. Rudant, and K. Belmkaddem, "Miniature directive antennas," *Int. J. Microw. Wirel. Technol.*, vol. 6, no. 1, pp. 45–50, Feb. 2014, doi: 10.1017/S1759078713001098.
[32] K. P. Lee and H. K. Choi, "Five-element ESPAR antenna using the annular ring slot active element," *Microw. Opt. Technol. Lett.*, vol. 58, no. 12, pp. 2800–2804, Dec. 2016, doi: 10.1002/MOP.30155.
[33] Q. Liang, B. Sun, and G. Zhou, "Multiple Beam Parasitic Array Radiator Antenna for 2.4 GHz WLAN Applications," *IEEE Antennas Wirel. Propag. Lett.*, vol. 17, no. 12, pp. 2513–2516, Dec. 2018, doi: 10.1109/LAWP.2018.2880208.
[34] C. A. Balanis, *Antenna Theory: Analysis and Design*. Wiley-Interscience; 3 edition, 2005.
[35] W. Geyi, "Physical limitations of antenna," *IEEE Trans. Antennas Propag.*, vol. 51, no. 8, pp. 2116–2123, Aug. 2003, doi: 10.1109/TAP.2003.814754.